\pdfoutput=1
\PassOptionsToPackage{expansion=false}{microtype} %
\documentclass[conf]{new-aiaa}
\ifdefined\pdfmapfile
  \pdfmapfile{+qtm.map} %
  \pdfmapfile{+newtx.map} %
\fi
\usepackage{graphicx}
\graphicspath{ {./figures/} }
\usepackage[rightcaption]{sidecap}
\usepackage{wrapfig}
\usepackage{float}
\usepackage{imakeidx}
\usepackage{comment}
\usepackage{commath}
\usepackage[titletoc]{appendix}
\usepackage{graphicx}
\usepackage{resizegather}
\usepackage[english]{babel}
\usepackage{tabu}
\usepackage{booktabs}
\usepackage{xfrac}
\usepackage{tabularx}
\usepackage{amsmath}
\usepackage{commath}
\usepackage{graphicx,bm}
\usepackage{verbatim}
\usepackage{lscape}
\usepackage{relsize}
\usepackage{enumitem}
\usepackage{textcomp}
\usepackage{breqn}
\usepackage{makecell}
\usepackage{longtable,tabularx}
\usepackage{multirow}
\usepackage{doi}
\usepackage{fancyhdr}
\usepackage{algorithm}
\usepackage{algpseudocode}
\usepackage[section]{placeins}
\algrenewcommand{\algorithmicrequire}{\textbf{Input:}}
\algrenewcommand{\algorithmicensure}{\textbf{Output:}}
\algrenewcommand{\algorithmiccomment}[1]{\hfill\makebox[0.34\linewidth][l]{\(\triangleright\)~#1}}
\usepackage{setspace}
\usepackage{footnote}
\PassOptionsToPackage{hyphens}{url}
\usepackage{hyperref}
\hypersetup{colorlinks,linkcolor={blue},citecolor={blue},urlcolor={blue}}
\usepackage[numbers]{natbib}
\usepackage{mathtools}
\usepackage[disable]{todonotes}
\usepackage[framemethod=tikz]{mdframed}
\usepackage{booktabs,xcolor,siunitx}
\usepackage{soul}
\usepackage{cleveref}

\newcommand{\f}[2]{\frac{#1}{#2}}
\newcommand{\mb}[1]{\mathbf{#1}}

\DeclareMathAlphabet\mathbfcal{OMS}{cmsy}{b}{n}
\renewcommand{\d}{\mathop{}\!\mathrm{d}} %
\newcommand{\p}{\partial}

\title{Matrix-Free Resolvent Analysis for \\Large-Scale Aerodynamic Systems}

\author{Rohit Sunil Kanchi\footnote{PhD Candidate, Department of Mechanical and Aerospace Engineering; AIAA Member; rkanchi@vols.utk.edu.}}
\author{Sicheng He\footnote{Assistant Professor, Department of Mechanical and Aerospace Engineering; AIAA Member; sicheng@utk.edu.}}
\affil{University of Tennessee, Knoxville, TN 37996}

\date{\today}


\begin{document}
\maketitle{}

\begin{abstract}
Predicting transonic buffet requires resolving both flow stability and disturbance amplification.
Resolvent analysis gives the amplification.
However, each gain needs the solution of tens of linear systems with the shifted Jacobian and its adjoint.
Sparse direct factorization, the usual route, is extrapolated to need more than $130$~TB for a three-dimensional jet with $4\times10^{7}$ unknowns.
For the stiff Reynolds-averaged operator of a transonic wing, iterative solution has been reported to stall on the adjoint system.
To avoid the direct factorization and to converge the adjoint, we present a matrix-free resolvent framework in the Reynolds-averaged Navier--Stokes solver ADflow that combines automatic differentiation, Krylov solution with subspace recycling, and Lanczos bidiagonalization, with backward-error checks on each singular triplet.
A three-level study of the NASA Common Research Model wing reaches $1.73\times10^{8}$ complex unknowns.
The coarsest grid predicts an unstable buffet mode, whereas no unstable eigenvalue is found on the finer grids within the shifts searched.
The finer-grid gain peaks are consistent with collective amplification by damped modes, whose responses retain the buffet-cell structure and travel outboard at approximately $0.3$ of the free-stream speed.
The peak gain remains sensitive to resolution despite these similar structures.
These findings motivate future mesh adaptation targeted at convergence of flow stability and disturbance amplification.
\end{abstract}

\section{Introduction}

Resolvent analysis of the Reynolds-averaged operators of transonic aircraft is limited by the linear solves with the shifted Jacobian and by the memory needed to assemble and factor it.
The resolvent gives the forcing that a linearized flow amplifies most and the response it produces~\cite{Farrell1993,Trefethen1993,schmid2007nonmodal}.
Its poles are the global eigenvalues that identify shock-buffet onset as a global instability of aerofoils, finite wings and elastic aircraft~\cite{Paladini2019b,Timme2020,HoutmanTimme2023}.
Onset can therefore be differentiated and constrained in design: by data-driven constraints~\cite{Li2022c,Li2023a}, by adjoint eigenvalue and eigenvector derivatives that have alleviated the buffet of the CRM wing~\cite{He2022a,He2023,Kanchi2026}, and by adjoints of limit-cycle stability and of the first Lyapunov coefficient~\cite{He2023a,He2026z}.
Resolvent analysis is reviewed by \citet{Taira2017,Taira2020} and \citet{Jovanovic2021}, and its compressible implementation by \citet{Rolandi2024}.

Each application of the resolvent is a linear solve with the shifted Jacobian.
Factorizing it~\cite{Amestoy2001,Bugeat2019,Ribeiro2020} becomes memory-bound in three dimensions: a $3.9\times10^{7}$-unknown jet would need more than $130$~TB~\cite{Farghadan2025}.
Time-stepping methods~\cite{MONOKROUSOS2010,Martini2021,Farghadan2025} need a stable or discounted operator and a long integration, and one-way formulations~\cite{Towne2022} exclude a shock.
Matrix-free frequency-domain methods take the direct and adjoint actions from the flow solver~\cite{FosasdePando2012,Ohmichi2021}; for the stiff RANS operator of a transonic wing the preconditioner has been the main difficulty~\cite{Vevek2023,Edwards2026}.
The closest work, the CRM resolvent of \citet{Houtman2023} in DLR-TAU~\cite{Houtman2022}, uses an assembled, hand-derived Jacobian and a power iteration, because its adjoint solves stalled and Lanczos bidiagonalization could not be converged~\cite{Vevek2023}.

Here the RANS Jacobian is never assembled.
Forward- and reverse-mode automatic differentiation (AD) in ADflow~\cite{ADflow2020} supply the shifted operator and its Hermitian transpose inside preconditioned Krylov solves, and thick-restart Lanczos bidiagonalization in SLEPc~\cite{slepc-user-ref} gives the singular triplets.
The framework extends the methods built on ADflow's differentiated residuals: time-spectral solvers and their adjoints~\cite{He2021c,He2020e,He2019b,He2022b}, torus time-spectral methods~\cite{He2024torus,KanchiTorusRANS}, a time-spectral resolvent~\cite{Howell2026}, transition prediction~\cite{Shi2020a} and a dataset of $5\times10^{5}$ aerofoil flows~\cite{Kanchi2025b}.
Potential applications extend beyond buffet to low-frequency shock/separation unsteadiness, inlet buzz and the unsteady aerodynamics of blunt-body entry vehicles.

The work makes three contributions.
The first is a matrix-free, certified frequency-domain resolvent of the RANS--Spalart--Allmaras equations, in which GCRO-DR recycles Krylov subspaces across the Lanczos bidiagonalization and the adjoint converges as the forward solve does.
The computational contribution is the integration of established components into this framework at RANS scale, rather than the individual algorithms (Fig.~\ref{fig:resolvent_overview}).
Its largest computation, $1.73\times10^{8}$ complex unknowns, is $2.3$ times the largest resolvent computation we know of~\cite{Houtman2022}.
The second is a set of composable norms including Chu's energy~\cite{Chu1965}, with the finding that the Euclidean and volume norms put the optimal forcing of the CRM into the turbulence variable.
The third is a three-level grid study of the CRM resolvent: the coarse level's unstable buffet mode gives way, on finer levels where no unstable eigenvalue is found, to amplification by damped modes of the same buffet-cell structure, and the gain peaks are read with the product $\xi\delta$, which tells a single-mode peak from a collective one.
The ingredients of $\xi\delta$ are established: the singular-value sensitivity~\cite{Brandt2011,FosasdePando2014}, the resolvent wavemaker~\cite{Qadri2017} and the wavemaker coefficient of \citet{Houtman2023}, the agreement of the gain times the eigenvalue distance with the inverse overlap when one eigenvalue dominates~\cite{Symon2018}, and the power law of a Jordan block~\cite{Moro1997,Trefethen2005}.
New here are the reading of the gain times the real part of the overlap as the sensitivity of the gain to a uniform damping, the bound $\xi\delta\le1$ for an isolated pole of any non-normality, and an estimate of $\xi\delta$ from three gains.

\section{Resolvent Formulation}
\label{sec:theory}

\begin{figure}[!t]
    \centering
    \includegraphics[width=\textwidth]{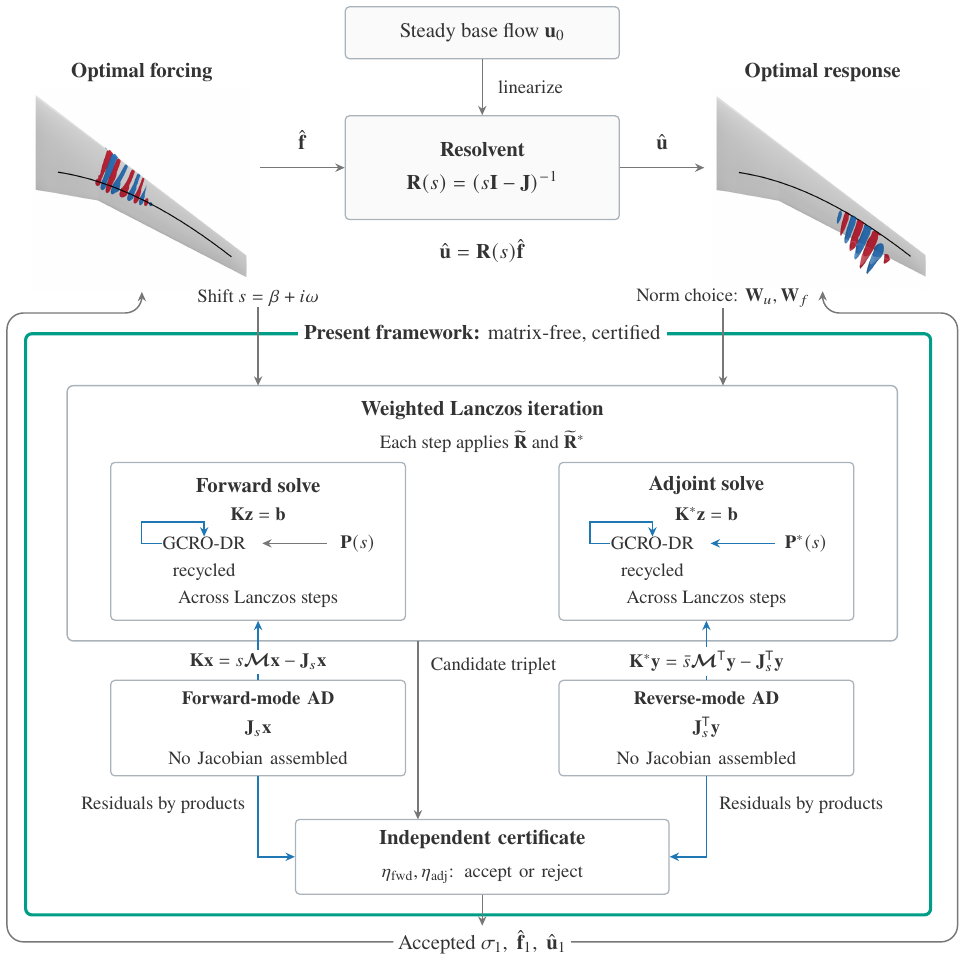}
    \caption{Resolvent map and present framework.
    The teal enclosure identifies the contribution: a combination of established components, with the couplings introduced here shown in blue.
    Each GCRO-DR solver recycles its own Krylov subspace across Lanczos steps; the blue $\mb{P}^{*}(s)$ link highlights the adjoint-preconditioning coupling.
    Certificate residuals are evaluated with operator products, independently of the inner-solver history.
    For the semi-definite Chu weight, full-response recovery requires one extra solve per direction (Section~\ref{sec:certification}).
    The L1 fields at $\omega=6.4$, $\beta=0$ show the optimal forcing and response, with amplitudes normalized separately.}
    \label{fig:resolvent_overview}
\end{figure}

The resolvent maps a forcing of the linearized flow to its response.
Figure~\ref{fig:resolvent_overview} connects this map to the matrix-free computation.
We first define this map, then the energy norm used to measure amplification, and finally the damping sensitivity used to interpret a gain peak.

\subsection{Linearized Dynamics and Resolvent}
\label{sec:resolvent}

ADflow~\cite{ADflow2020} discretizes the compressible RANS equations with the Spalart--Allmaras (SA) model~\cite{Spalart1992} by a second-order cell-centred finite-volume method, with the state $\mb{u}=[\rho,u_1,u_2,u_3,\rho E,\tilde\nu]^\intercal$ per cell, $n=6N_{\mathrm{cells}}$.
The linearization keeps $\tilde\nu$ as a state variable, as the DLR-TAU buffet computations do~\cite{He2020,Houtman2023}.
Linearizing about a steady state $\mb{u}_0$, $\mb{r}(\mb{u}_0)=\mb{0}$, with a forcing $\mb{f}(t)$ gives
\begin{equation}
\mathbfcal{M}\f{\partial \tilde{\mb{u}}}{\partial t}
=\mb{J}_s\tilde{\mb{u}}+\mathbfcal{M}\mb{f}(t).
\label{eq:linearized}
\end{equation}
Here the mass matrix $\mathbfcal{M}\in\mathbb{R}^{n\times n}$ and spatial Jacobian $\mb{J}_s\in\mathbb{R}^{n\times n}$ are defined by
\begin{equation*}
\mathbfcal{M}\coloneqq\left.\f{\p\mb{w}_c}{\p\mb{u}}\right|_{\mb{u}_0},
\qquad
\mb{J}_s\coloneqq\left.\f{\p\mb{r}}{\p\mb{u}}\right|_{\mb{u}_0},
\end{equation*}
where $\mb{w}_c$ is the conservative state and $\mb{r}$ is the spatial residual.

For perturbations $\tilde{\mb{u}}=\hat{\mb{u}}e^{st}$ and forcing $\mb{f}=\hat{\mb{f}}e^{st}$, Eq.~\eqref{eq:linearized} defines the response $\hat{\mb{u}}=\mb{R}(s)\hat{\mb{f}}$ through the resolvent
\begin{equation}
\mb{R}(s)=(s\mb{I}-\mb{J})^{-1}.
\label{eq:resolvent}
\end{equation}
Here $\mb{J}\coloneqq\mathbfcal{M}^{-1}\mb{J}_s$ is the evolution operator.
The classical resolvent, $s=i\omega$, requires a stable $\mb{J}$; for an unstable operator the contour is shifted to $s=\beta+i\omega$ with $\beta>\max_k\operatorname{Re}\lambda_k$, which gives the finite-time-horizon gain~\cite{Jovanovic2004,Yeh2019}.

\subsection{Energy Norm and Gain}
\label{sec:norm_def}

Hermitian positive semi-definite weights $\mb{W}_u,\mb{W}_f\in\mathbb{C}^{n\times n}$ measure the response and forcing, with $\|\mb{a}\|_W^2\coloneqq\mb{a}^{*}\mb{W}\mb{a}$.
When the forcing weight is singular, the forcing is restricted to its range.
The leading gain is the largest ratio of response norm to forcing norm,
\begin{equation}
\sigma_1(s)=\max_{\substack{\hat{\mb{f}}\in\operatorname{range}(\mb{W}_f)\\\hat{\mb{f}}\neq\mb{0}}}
\f{\|\mb{R}(s)\hat{\mb{f}}\|_{\mb{W}_u}}{\|\hat{\mb{f}}\|_{\mb{W}_f}}.
\label{eq:gain}
\end{equation}
This optimization determines the optimal forcing and its amplified response.

To compute the gain in Eq.~\eqref{eq:gain}, factor each weight as $\mb{W}=\mb{F}^{2}$ with $\mb{F}$ Hermitian positive semi-definite, and denote its pseudo-inverse by $\mb{F}^{+}$.
The corresponding weighted resolvent is
\begin{equation}
\tilde{\mb{R}}(s)=\mb{F}_u\,\mb{R}(s)\,\mb{F}_f^{+}.
\label{eq:weighted_resolvent}
\end{equation}
Its leading singular value is $\sigma_1(s)$, so the weighted amplification problem becomes a singular value decomposition.

A weight is cellwise, $\mb{W}=\operatorname{blkdiag}(V_i\mb{H}_i)$: the cell volume as quadrature composed with a physical metric $\mb{H}_i$.
The metric used on the wing follows Chu's compressible disturbance energy~\cite{Chu1965},
\begin{equation}
E_{\mathrm{Chu}}=\f12\int_\Omega\Big[\f{R\bar T}{\bar\rho}\,|\rho'|^2+\bar\rho\,|\mb{u}'|^2+\f{\bar\rho c_v}{\bar T}\,|T'|^2\Big]\d V.
\label{eq:chu}
\end{equation}
On ADflow's state this gives a dense $6\times6$ block per cell, of rank five, with no energy in $\tilde\nu$.

\subsection{Damping Sensitivity and the Index $\xi\delta$}
\label{sec:wavemaker_def}

Near an eigenvalue $\lambda$, the leading gain has the pole approximation
\begin{equation}
\sigma_1(s)\approx\f{\kappa_\lambda}{|s-\lambda|}.
\label{eq:pole}
\end{equation}
Here $\kappa_\lambda\coloneqq\|\mb{q}\|\,\|\mb{p}\|/|\mb{p}^{*}\mb{q}|$ is the eigenvalue condition number, with $\mb{q}$ and $\mb{p}$ the right and left eigenvectors.
This single-mode approximation provides the reference for interpreting a gain peak.

With the same weight on forcing and response, define the inner product $\langle\mb{a},\mb{b}\rangle_W\coloneqq\mb{a}^{*}\mb{W}\mb{b}$.
A perturbation $\delta\mb{J}$ of the operator changes the leading gain by~\cite{Brandt2011,FosasdePando2014}
\begin{equation}
\delta\ln\sigma_1=\sigma_1\operatorname{Re}\langle\mb{f}_1,\delta\mb{J}\,\mb{u}_1\rangle_W.
\label{eq:gain_sensitivity}
\end{equation}
Restricted to one cell it is the resolvent wavemaker~\cite{Qadri2017,Skene2022}.

For the uniform perturbation $\delta\mb{J}=-\beta\mb{I}$, the contour shift defines the damping sensitivity
\begin{equation}
\begin{aligned}
\xi&\coloneqq-\f{\p\ln\sigma_1}{\p\beta}\\[3pt]
&=\sigma_1\operatorname{Re}\langle\mb{f}_1,\mb{u}_1\rangle_W.
\end{aligned}
\label{eq:xi}
\end{equation}
It measures the fractional rise of the gain per unit of damping removed from every mode; \citet{Houtman2023} used the modulus of the overlap expression as a wavemaker coefficient.
The second equality holds for a definite weight.
For the semi-definite Chu weight it omits the response of the unobserved turbulence variable, so the overlap gives an estimate of $\xi$ and the central difference of $\ln\sigma_1$ in $\beta$ gives $\xi$ itself.

Let $\delta$ be the distance from $s$ to the nearest eigenvalue and $\psi\coloneqq\arg(s-\lambda)$.
In the isolated-pole limit of Eq.~\eqref{eq:pole}, $\xi\delta=\cos\psi\le1$ however non-normal the mode is, whereas the Jordan-block limit of order $m$ gives $m\cos\psi$.
Both values are limits; between them $\xi\delta$ is a diagnostic of how collective a peak is, not a count of modes: about one for the resonance of a single mode, well above one for a pseudo-resonance.

Without an eigenvalue, three gains at the peak frequency give the estimates
\begin{equation}
\begin{aligned}
\delta_{\mathrm{res}}&\coloneqq-\f{\p_\beta\ln\sigma_1}{\p_\beta^{2}\ln\sigma_1},\\[3pt]
m_\beta&\coloneqq\xi\delta_{\mathrm{res}}.
\end{aligned}
\label{eq:gain_only_index}
\end{equation}
The quantities in Eq.~\eqref{eq:gain_only_index} estimate $\delta$ and $\xi\delta$, respectively, and are exact in the two limits.

\section{Matrix-Free Computation}
\label{sec:matrix_free}

Each application of the resolvent is computed through a linear solve.
Defining the shifted operator $\mb{K}(s)\coloneqq s\mathbfcal{M}-\mb{J}_s$, the response to a forcing $\hat{\mb{f}}$ satisfies
\begin{equation}
\mb{K}(s)\hat{\mb{u}}=\mathbfcal{M}\hat{\mb{f}}.
\label{eq:resolvent_solve}
\end{equation}
This evaluates the map in Eq.~\eqref{eq:resolvent} without forming the resolvent matrix.

\subsection{Operator and Singular Value Decomposition}
\label{sec:mf_operator}

Thick-restart Lanczos bidiagonalization~\cite{Baglama2005,Hernandez2008} in SLEPc computes the leading triplets of $\tilde{\mb{R}}$.
Each step applies $\tilde{\mb{R}}=\mb{F}_u\mb{K}^{-1}\mathbfcal{M}\mb{F}_f^{+}$ and its adjoint once, that is, one solve of Eq.~\eqref{eq:resolvent_solve} and one with $\mb{K}^{*}=\bar{s}\mathbfcal{M}^{\intercal}-\mb{J}_s^{\intercal}$.
The products $\mb{J}_s\mb{x}$ and $\mb{J}_s^{\intercal}\mb{y}$ are the forward- and reverse-mode AD products of ADflow's residual, each costing about one residual evaluation and including the turbulence equation.
On the CRM they agree with the assembled Jacobian to $1.6\times10^{-14}$.
Every matrix, vector, Krylov solver and the SVD are complex, through complex builds of PETSc~\cite{petsc-user-ref} and SLEPc.
Unlike a power iteration, whose rate is $\sigma_2/\sigma_1$, Lanczos bidiagonalization converges $\sigma_1$ in the same number of applications on and off resonance, 22 on the CRM.

\subsection{Preconditioning, Recycling and the Adjoint Solve}
\label{sec:preconditioning}

The solves with $\mb{K}$ and $\mb{K}^{*}$ are preconditioned from
\begin{equation}
\mb{P}(s)=s\mathbfcal{M}-\tilde{\mb{J}}_s,
\label{eq:pmat}
\end{equation}
where $\tilde{\mb{J}}_s$ is the first-order Jacobian of ADflow's Newton--Krylov and adjoint solvers~\cite{Yildirim2019}, with about a fifth of the nonzeros of $\mb{J}_s$.
The preconditioner is ADflow's adjoint preconditioner, one-level additive Schwarz with ILU(1) in reverse Cuthill--McKee order per subdomain, applied three times per Krylov iteration.
The Krylov method is GCRO-DR~\cite{Parks2006} through PETSc's HPDDM interface~\cite{Jolivet2021}, with $16$ recycled vectors kept across the right-hand sides of the Lanczos recurrence.
Recycling needs $\mb{C}=\mb{K}\mb{U}$ for the operator actually applied, which the AD product, linear in the perturbation, satisfies exactly.
The adjoint has its own Krylov solver, preconditioned from $\mb{P}^{*}$, which ADflow assembles by transposed insertion; it converges in as many iterations as the forward solve.

\subsection{Certification by Backward Error}
\label{sec:certification}

SLEPc measures convergence against the operator it is handed, here an inexact inner solve, and a wrong adjoint yields a converged triplet of the wrong operator.
Every triplet is therefore checked against the operators.
With $\mb{f}=\mb{F}_f^{+}\mb{v}$ and $\mb{q}=\mb{F}_u^{+}\mb{u}$,
\begin{equation}
\begin{aligned}
\mathbfcal{M}\mb{f}&=\sigma\mb{K}\mb{q},\\[3pt]
\mb{W}_u\mb{q}&=\sigma\,\mb{K}^{*}\mathbfcal{M}^{-\intercal}\mb{W}_f\mb{f},
\end{aligned}
\label{eq:triplet}
\end{equation}
which contain only operator products and so hold however the inner solve behaved.
Writing $\mb{t}\coloneqq\mathbfcal{M}^{-\intercal}\mb{W}_f\mb{f}$, the backward errors checked in the forward and adjoint directions are
\begin{equation}
\begin{aligned}
\eta_{\mathrm{fwd}}&=\f{\|\mathbfcal{M}\mb{f}-\sigma\mb{K}\mb{q}\|}{\|\mathbfcal{M}\|\,\|\mb{f}\|+\sigma\|\mb{K}\|\,\|\mb{q}\|},\\[6pt]
\eta_{\mathrm{adj}}&=\f{\|\mb{W}_u\mb{q}-\sigma\mb{K}^{*}\mb{t}\|}{\|\mb{W}_u\mb{q}\|+\sigma\|\mb{K}\|\,\|\mb{t}\|}.
\end{aligned}
\label{eq:backward}
\end{equation}
The operator norms are infinity norms; a frequency is accepted only if both errors pass the gate $100\,\epsilon_{\mathrm{svd}}$.
On the matrix-free path $\mb{K}$ has no matrix norm and $\|\mb{K}\|$ is replaced by $\|\mb{P}\|$, so $\eta$ is a backward error relative to that normalization.
For the semi-definite Chu weight the full response $\mb{q}=\mb{K}^{-1}\mathbfcal{M}\mb{f}$ is computed by one more solve in each direction and its backward error gated.

\subsection{Summary of the Algorithm}
\label{sec:algorithm}

Algorithm~\ref{alg:resolvent} summarizes the computation at one shift $s$; a frequency sweep repeats it at $s=\beta+i\omega$ with fixed $\beta$.
The forward and adjoint solvers $\mathcal{S}_f$ and $\mathcal{S}_a$ keep separate recycled subspaces, $\eta_{\mathrm{svd}}$ is SLEPc's residual estimate, and $\mathrm{ok}$ records that every inner solve passed.

\begin{algorithm}[!htbp]
\caption{Matrix-free resolvent computation at one frequency.}
\label{alg:resolvent}
\begin{algorithmic}[1]
\Require $\mb{u}_0$, $s$, $\mb{W}_u$, $\mb{W}_f$, $\epsilon_{\mathrm{lin}}$, $\epsilon_{\mathrm{svd}}$, $n_{\mathrm{cv}}$
\Ensure $(\sigma_1,\mb{f}_1,\mb{q}_1)$ or $\varnothing$
\Statex \textbf{Operator setup}
\State $(\mathbfcal{M},\tilde{\mb{J}}_s,\tilde{\mb{J}}_s^{\intercal})\gets\textsc{Setup}(\mb{u}_0)$ \Comment{mass and first-order Jacobian}
\State $\mb{K}\mb{x}\coloneqq s\mathbfcal{M}\mb{x}-\mb{J}_s\mb{x}$ \Comment{forward AD}
\State $\mb{K}^{*}\mb{y}\coloneqq\bar{s}\mathbfcal{M}^{\intercal}\mb{y}-\mb{J}_s^{\intercal}\mb{y}$ \Comment{reverse AD}
\State $\mb{P}\gets s\mathbfcal{M}-\tilde{\mb{J}}_s$ \Comment{forward preconditioner matrix}
\State $\mb{P}^{*}\gets\bar{s}\mathbfcal{M}^{\intercal}-\tilde{\mb{J}}_s^{\intercal}$ \Comment{adjoint preconditioner matrix}
\State $(\mb{F}_u,\mb{F}_u^{+},\mb{F}_f,\mb{F}_f^{+})\gets\textsc{WeightFactors}(\mb{W}_u,\mb{W}_f)$ \Comment{Section~\ref{sec:norm_def}}
\Statex \smallskip\textbf{Singular-triplet computation}
\State $\mathcal{S}_f\gets\mathrm{GCRO\text{-}DR}(\mb{K},\mb{P};\epsilon_{\mathrm{lin}})$ \Comment{forward recycling space}
\State $\mathcal{S}_a\gets\mathrm{GCRO\text{-}DR}(\mb{K}^{*},\mb{P}^{*};\epsilon_{\mathrm{lin}})$ \Comment{adjoint recycling space}
\State $\tilde{\mb{R}}\mb{x}\coloneqq\mb{F}_u\,\mathcal{S}_f(\mathbfcal{M}\mb{F}_f^{+}\mb{x})$ \Comment{one forward solve per application}
\State $\tilde{\mb{R}}^{*}\mb{y}\coloneqq\mb{F}_f^{+}\mathbfcal{M}^{\intercal}\mathcal{S}_a(\mb{F}_u\mb{y})$ \Comment{one adjoint solve per application}
\State $(\sigma_1,\mb{u}_1,\mb{v}_1,\eta_{\mathrm{svd}})\gets\textsc{TRLanczos}(\tilde{\mb{R}},\tilde{\mb{R}}^{*};\epsilon_{\mathrm{svd}},n_{\mathrm{cv}})$ \Comment{Section~\ref{sec:mf_operator}}
\Statex \smallskip\textbf{Verification and acceptance}
\State $(\mb{f}_1,\mb{q}_1,\eta_{\mathrm{fwd}},\eta_{\mathrm{adj}})\gets\textsc{CheckTriplet}(\sigma_1,\mb{u}_1,\mb{v}_1)$ \Comment{Section~\ref{sec:certification}}
\If{$\mathrm{ok}\ \land\ \eta_{\mathrm{svd}}\le\epsilon_{\mathrm{svd}}\ \land\ \max(\eta_{\mathrm{fwd}},\eta_{\mathrm{adj}})\le100\,\epsilon_{\mathrm{svd}}$}
  \State \Return $(\sigma_1,\mb{f}_1,\mb{q}_1)$
\Else
  \State \Return $\varnothing$ \Comment{recorded as missing}
\EndIf
\end{algorithmic}
\end{algorithm}

\section{Verification on a Laminar Cylinder}
\label{sec:cylinder}

\begin{figure}[!htbp]
    \centering
    \includegraphics[width=0.55\textwidth]{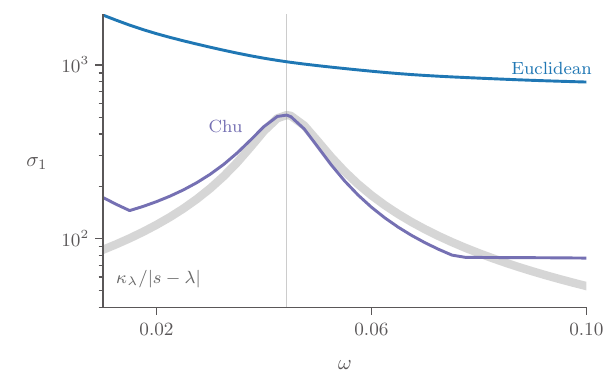}\\[4pt]
    \includegraphics[width=0.80\textwidth]{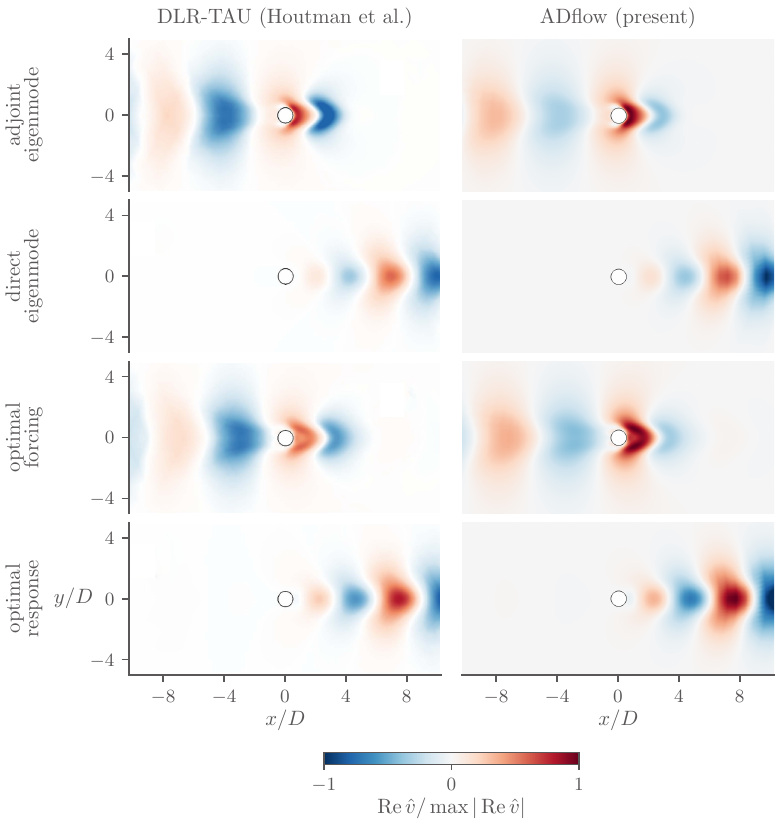}
    \caption{The cylinder.
    Top: leading gain at $M=0.05$, $Re=70$, $\beta=0.01$ in the Euclidean and volume-weighted Chu norms, with the pole prediction $\kappa_\lambda/|s-\lambda|$ fitted to the Chu curve (thick gray); the vertical line is the eigenfrequency.
    Bottom: real part of the cross-stream velocity at $Re=40$, $M=0.2$ of the adjoint and direct eigenmodes and of the optimal forcing and response at $\omega=0.7$, from DLR-TAU (left, \citet[Fig.~10(c--f)]{Houtman2023}, reproduced under the CC BY 4.0 licence) and ADflow (right, each field scaled by its maximum and phase-aligned with the reference); the colour bar applies to the present fields.}
    \label{fig:cyl}
\end{figure}

On a laminar cylinder at $M=0.05$ and $Re=70$ ($8192$ cells), shift-invert Krylov--Schur gives the unstable shedding eigenvalue $\lambda=4.197\times10^{-3}+4.421\times10^{-2}\,i$.
Walking $\beta$ toward $\operatorname{Re}\lambda$ at $\omega=\operatorname{Im}\lambda$, a quadratic fit of $1/\sigma_1$ recovers $\operatorname{Re}\lambda$ to $1.6\times10^{-11}$, and twenty iterative solver configurations give the same $\sigma_1$ to eleven digits.
In the Chu norm the gain follows the pole term of Eq.~\eqref{eq:pole} with $\kappa_\lambda=2.98$ (Fig.~\ref{fig:cyl}, top).
\citet{Houtman2023} verified DLR-TAU on a laminar cylinder at $Re=40$, $M=0.2$.
Recomputed on a $9\,984$-cell O-grid with the same far field and inner product, the eigenvalue agrees to $1.0\%$ in growth rate and $0.2\%$ in frequency (Table~\ref{tab:cyl_houtman}), and the eigenmodes and optimal modes reproduce theirs (Fig.~\ref{fig:cyl}, bottom).
The gain near this damped wake mode is not grid-converged in either code: it grows with the length of wake a grid keeps.

\begin{table}[!htbp]
\centering
\caption{The DLR-TAU cylinder of \citet[Appendix~B]{Houtman2023} at $Re=40$, $M=0.2$ and $\omega=0.7$, recomputed on a grid of the same size and far-field distance in the volume-weighted inner product of the conservative variables.
Eigenvalues are in $D/U_\infty$; the present gains are given in ADflow's time unit $D/\sqrt{p_\infty/\rho_\infty}$ and in $D/U_\infty$, since the publication does not state the unit of the reference gains.}
\label{tab:cyl_houtman}
\small
\begin{tabular}{lccc}
\toprule
 & \citet{Houtman2023} & present, $D/\sqrt{p_\infty/\rho_\infty}$ & present, $D/U_\infty$ \\
\midrule
$\lambda_1$ & $-0.0275+0.7145\,i$ & \multicolumn{2}{c}{$-0.02776+0.71595\,i$} \\
$\sigma_1$ & $2158.98$ & $2478.2$ & $586.4$ \\
$\sigma_2$ & $162.66$ & $235.1$ & $55.6$ \\
\bottomrule
\end{tabular}
\end{table}

\section{Shock-Buffet Amplification on the CRM Wing}
\label{sec:results}

\subsection{Configuration}
\label{sec:crm_case}

The configuration is the wing of the NASA Common Research Model (CRM)~\cite{Vassberg2008}, without fuselage and tail, at $M=0.85$, $Re=5\times10^{6}$ on the mean aerodynamic chord $\bar c=1$~m and $\alpha=4.2^\circ$.
The three grid levels are uniform $r=2$ coarsenings of one structured grid (Table~\ref{tab:crm3M_grids}), whose lifts converge at order $2.24$~\cite{Kanchi2026}; coarsening mainly coarsens the boundary layer.
The base flows are converged to relative residuals of $1.6\times10^{-12}$, $8.2\times10^{-12}$ and $9.9\times10^{-11}$ on L2, L1 and L0.
L0 needs $724$ nonlinear iterations against $205$ on L1 and stops at a higher residual; whether this changes its eigenvalues or gain is not tested here.
Frequencies and gains are in ADflow's units, with time unit $\bar c/\sqrt{p_\infty/\rho_\infty}$; since $M\sqrt{\gamma}=1.006$ they are within $1\%$ of chord-based values, and $\omega=3.741$ is $\mathrm{St}=0.59$, or $175$~Hz.
The norm is the volume-weighted Chu energy; on L2 three others are compared.

\begin{table}[!htbp]
\centering
\caption{Grid levels of the wing-only CRM at $M=0.85$, $Re=5\times10^{6}$, $\alpha=4.2^\circ$, after the companion grid study~\cite{Kanchi2026}; $n$ is the number of unknowns.}
\label{tab:crm3M_grids}
\small
\begin{tabular}{llrrr}
\toprule
level & cells & $n$ & $C_L$ & $C_D$ [counts] \\
\midrule
L0 & $28\,835\,840$ & $173\,015\,040$ & $0.7031$ & $460.7$ \\
L1 & $3\,604\,480$ & $21\,626\,880$ & $0.6990$ & $463.5$ \\
L2 & $450\,560$ & $2\,703\,360$ & $0.6796$ & $464.2$ \\
\bottomrule
\end{tabular}
\end{table}

\subsection{Stability}
\label{sec:stability}

\begin{figure}[!htbp]
    \centering
    \includegraphics[width=\textwidth]{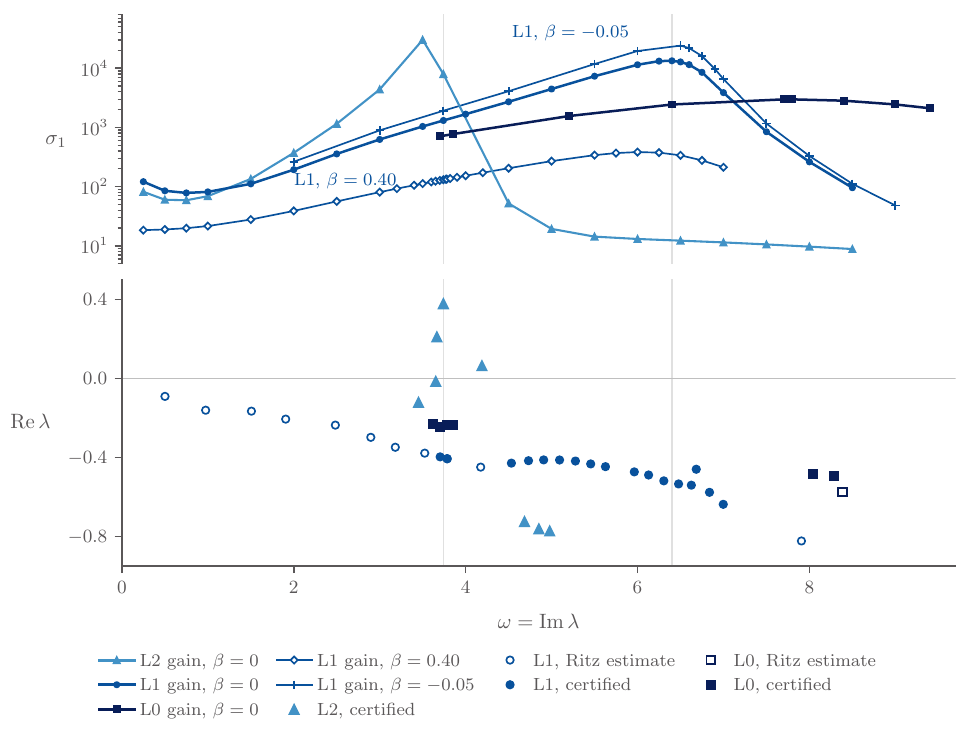}
    \caption{Gain and spectrum on a common frequency axis. Top: leading gain in the volume-weighted Chu norm on the imaginary axis, $\beta=0$, for L2, L1 and L0, and for L1 also on the contours $\beta=0.40$ and $-0.05$ (open diamonds and plus signs); mesh levels are blue, light to dark from L2 to L0; the L2 operator is unstable, so its curve is not a realizable response. Bottom: certified eigenvalues of L2 (triangles), L1 (filled circles) and L0 (filled squares), and unconverged Ritz estimates (open symbols). Vertical lines: L2 buffet frequency and L1 gain maximum.}
    \label{fig:crm3M_eigs}
\end{figure}

L2 has an unstable buffet mode, $\lambda_1=0.378+3.741\,i$, and two further unstable modes; an unforced time-accurate simulation on L2 grows at its frequency at the rate $0.389$, against $0.372$ predicted with the damping of the time scheme.
On L2 the gain is therefore taken on the contour $\beta=0.40$.
On L1 shift-invert Krylov--Schur certifies $17$ eigenvalues between $\omega=3.7$ and $7.0$, all with $\operatorname{Re}\lambda\le-0.40$ (Table~\ref{tab:crm3M_eigs}), and a seeded simulation and a rigid pitch pulse both decay.
The shifts searched are the three unstable L2 eigenvalues, $0.10+i\omega$ for $\omega\le4.2$ and $6.4\,i$ on L1, and $3.75\,i$ and $8.3\,i$ on L0; no unstable eigenvalue is found near any of them, but one elsewhere is not excluded.
The L0 eigenvalues found are damped too, but its buffet band lies at $\operatorname{Re}\lambda=-0.23$ to $-0.25$, against $-0.40$ on L1.
That refinement moves the operator before onset therefore means only that no unstable eigenvalue is found on L1 and L0; the damping of the buffet band is not converged with the grid.
This is linear evidence: it does not exclude a self-sustained state reached at finite amplitude.
On L1 and L0 the gain is taken on the imaginary axis, where it is the amplification of a harmonic forcing (Fig.~\ref{fig:crm3M_eigs}).

\begin{table}[!htbp]
\centering
\caption{Certified eigenvalues of the L1 operator at $\alpha=4.2^\circ$ from shift-invert Krylov--Schur on the resolvent shell, each pair certified by its backward error against $\mb{K}$ and $\mb{M}$.
The Strouhal number uses the reference chord and $U_\infty$.
An eigenvalue found from two shifts agrees to $10^{-3}$, which is the practical accuracy of these non-normal eigenvalues despite the backward errors.}
\label{tab:crm3M_eigs}
\small
\begin{tabular}{rrrl@{\hspace{2em}}rrrl}
\toprule
$\operatorname{Re}\lambda$ & $\operatorname{Im}\lambda$ & St & bwd.\ error & $\operatorname{Re}\lambda$ & $\operatorname{Im}\lambda$ & St & bwd.\ error \\
\midrule
$-0.3974$ & $3.7036$ & $0.586$ & $7\times10^{-11}$ & $-0.4736$ & $5.9625$ & $0.944$ & $9\times10^{-12}$ \\
$-0.4064$ & $3.7846$ & $0.599$ & $6\times10^{-11}$ & $-0.4893$ & $6.1291$ & $0.970$ & $1\times10^{-11}$ \\
$-0.4293$ & $4.5327$ & $0.717$ & $1\times10^{-12}$ & $-0.5189$ & $6.3057$ & $0.998$ & $2\times10^{-11}$ \\
$-0.4169$ & $4.7315$ & $0.749$ & $2\times10^{-11}$ & $-0.5347$ & $6.4778$ & $1.025$ & $6\times10^{-11}$ \\
$-0.4129$ & $4.9078$ & $0.777$ & $6\times10^{-12}$ & $-0.5409$ & $6.6263$ & $1.049$ & $5\times10^{-12}$ \\
$-0.4133$ & $5.0931$ & $0.806$ & $2\times10^{-12}$ & $-0.4602$ & $6.6834$ & $1.058$ & $2\times10^{-12}$ \\
$-0.4185$ & $5.2775$ & $0.835$ & $7\times10^{-12}$ & $-0.5770$ & $6.8376$ & $1.082$ & $4\times10^{-12}$ \\
$-0.4332$ & $5.4553$ & $0.863$ & $2\times10^{-12}$ & $-0.6378$ & $6.9977$ & $1.108$ & $3\times10^{-12}$ \\
$-0.4471$ & $5.6271$ & $0.891$ & $7\times10^{-12}$ & & & & \\
\bottomrule
\end{tabular}
\end{table}

\subsection{Gain and Norm on the Coarse Level}
\label{sec:crm_results}

\begin{figure}[!htbp]
    \centering
    \includegraphics[width=\textwidth]{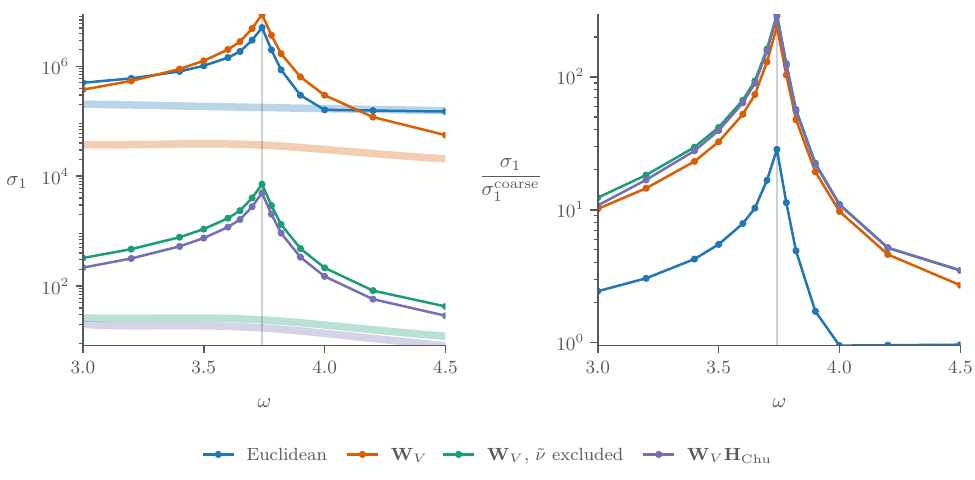}
    \caption{Leading gain on L2 across the buffet band at $\beta=0.40$. Left: $\sigma_1(\omega)$ on L2 (thin, markers) and on its one-level coarsening (thick, transparent) in the Euclidean norm, the volume norm $\mb{W}_V$, the volume norm without the turbulence variable, and the volume-weighted Chu norm; the vertical line is the buffet eigenfrequency. Right: ratio of the L2 gain to the coarse gain.}
    \label{fig:crm_sigma}
\end{figure}

On L2 all four norms peak at the buffet eigenfrequency (Fig.~\ref{fig:crm_sigma}); the Euclidean peak, $5.07\times10^{6}$, matches the pole term with $|s-\lambda_1|=0.022$.
In the Euclidean and volume norms the optimal forcing holds $100\%$ of its energy in $\tilde\nu$.
This is an artefact of units: $\tilde\nu$ is $\mathcal{O}(10^{-4})$ in ADflow's units, so a unit-norm forcing of its equation is a relative eddy-viscosity perturbation four orders of magnitude larger than a velocity one.
Excluding $\tilde\nu$ lowers $\sigma_1$ at the buffet frequency from $8.82\times10^{6}$ to $7.14\times10^{3}$; the Chu norm gives $4.95\times10^{3}$, with its forcing at the shock foot.
On a one-level coarsening of L2 the Euclidean gain off resonance agrees to $5\%$, but the physical norms differ by factors of $2.7$ to $5.2$ and the coarse mesh has no unstable mode: the growth rate of the buffet mode is not converged on L2.

\subsection{Gain on the Fine Levels}
\label{sec:crm3M}

On L1 (Fig.~\ref{fig:crm3M_eigs}, top) the $24$ certified points rise from $78$ near $\omega=0.75$ to a peak of $1.33\times10^{4}$ at $\omega=6.4$ and fall to $96$ by $\omega=8.5$; with $\sigma_2=43.4$ the peak is strongly rank one.
It is not a buffet resonance: on the contour $\beta=0.40$, L1 returns $130$ at the L2 buffet frequency, where L2 gives $4.95\times10^{3}$, and peaks out of the band, at $\omega=6.0$.
The L1 maximum moves with the contour, from $\omega=6.0$ at $\beta=0.40$ to $6.4$ at $\beta=0$ and $6.5$ at $\beta=-0.05$, which one isolated pole cannot do.
A discount inherited from the coarse mesh would understate the gain about thirtyfold and misplace its maximum.

L0, with $28.8$ million cells, overruns the 32-bit index of the assembled Jacobian and runs matrix-free only.
At $\omega=6.4$ its gain is $2.43\times10^{3}$, a factor of $5.5$ below L1, while $\sigma_2=43.1$ is unchanged.
Its maximum lies near $\omega=7.7$ with $2.97\times10^{3}$, on a broad curve that varies by a factor of $1.9$ over $5.2\le\omega\le9.4$.
At $\omega=6.4$ only $\sigma_2$ agrees between L1 and L0, to $1\%$; the peak gain falls $5.5$-fold and its frequency moves from $6.4$ to $7.7$.
Converged steady forces and similar mode shapes therefore do not imply converged amplification: the gain of a pseudo-resonance changes exponentially with the damping of its band (Section~\ref{sec:wavemaker}).

\subsection{Optimal Modes and Buffet Cells}
\label{sec:modes}

\begin{figure}[!htbp]
    \centering
    \includegraphics[width=0.92\textwidth]{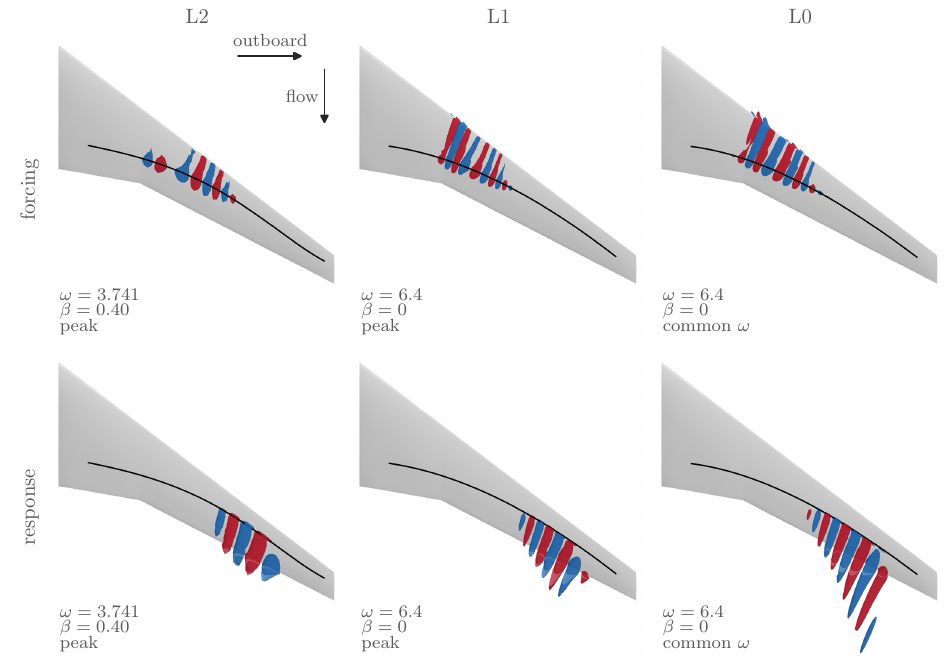}
    \caption{Optimal forcing (top) and response (bottom) in the volume-weighted Chu norm on L2, L1 and L0 (left to right), seen from above with the flow from top to bottom.
    Iso-surfaces of $\operatorname{Re}u'$ at $\pm25\%$ of its $99.9$th percentile (red positive, blue negative), the wing (gray) and the shock line (black); crop and scale are identical in every panel.
    Each panel gives its frequency and contour; L2 and L1 are at their gain peaks and L0 at the common frequency $\omega=6.4$, below its peak at $7.7$.}
    \label{fig:crm_response}
\end{figure}

\begin{figure}[!htbp]
    \centering
    \includegraphics[width=0.92\textwidth]{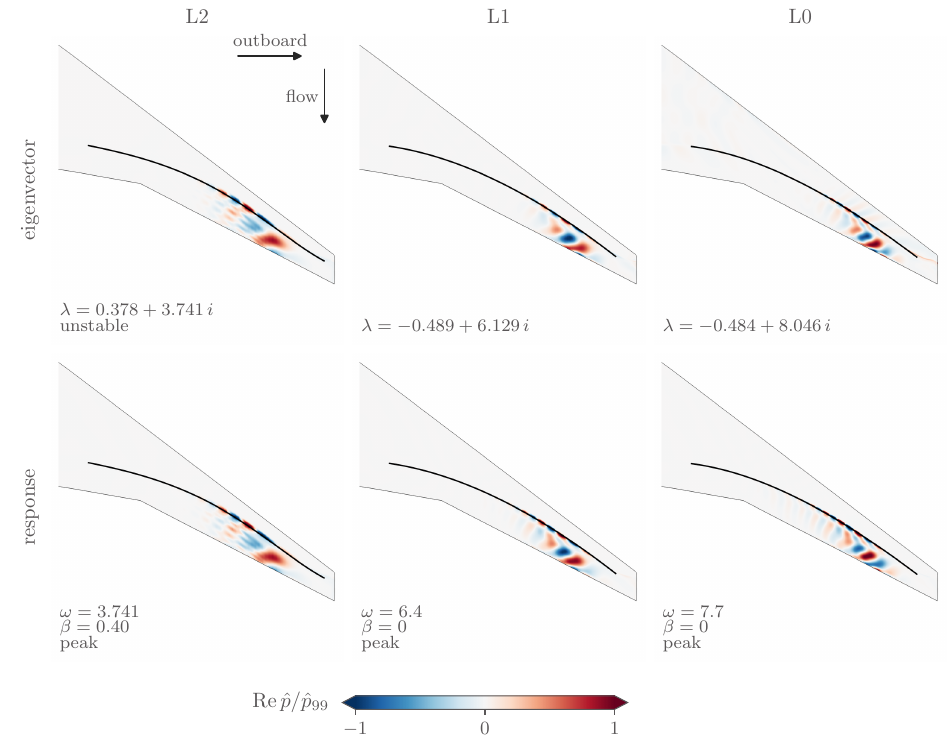}
    \caption{Upper-surface footprint of the buffet cells, $\operatorname{Re}\hat p/\hat p_{99}$, the wall-pressure perturbation normalized by its 99th percentile, on L2, L1 and L0 (left to right), seen from above.
    Top: the L2 unstable eigenvector and the certified L1 and L0 eigenvectors beneath the gain peaks, with eigenvalues as labelled.
    Bottom: the response of each level at its own gain peak, $\omega=3.741$, $6.4$ and $7.7$.
    The arrows give the free-stream direction (flow) and the spanwise direction toward the tip (outboard), against which the phase travel of the cells is measured; the black line is the shock.}
    \label{fig:crm3M_cells}
\end{figure}

On every level the optimal response is a row of alternating streamwise-velocity cells aft of the shock on the outboard wing, and the forcing a row of narrower cells further inboard, at and ahead of the shock (Fig.~\ref{fig:crm_response}).
On L2, $0.022$ from the pole, the response is the buffet eigenmode.
The direction of travel follows from the phase of the complex wall pressure $\hat p$ on the upper surface, not from a single snapshot: the spanwise and chordwise wavenumbers are amplitude-weighted mean phase increments, a phase that falls outboard means outboard travel, and downstream travel along the chord fixes the sign (Fig.~\ref{fig:crm3M_cells}, Table~\ref{tab:crm3M_cells}).
The L2 eigenvector cells, at $\omega=3.741$, are $0.61$ local chords long and travel outboard at $0.24\,U_\infty$; the L1 response cells, at $\omega=6.4$, are $0.42$ chords long at $0.28\,U_\infty$.
At the same $\omega=6.4$ the L0 cells are $0.43$ chords long at $0.30\,U_\infty$: the two finest levels agree in cell length to $2\%$ and in speed to $7\%$, while their gains differ $5.5$-fold.
At the L0 maximum, $\omega=7.7$, the cells shorten to $0.35$ chords at $0.29\,U_\infty$.
These are the finite-wing buffet cells of \citet{Timme2020} and \citet{Houtman2023}.
Since the frequency is the spanwise wavenumber times the phase speed, the shorter fine-level cells account for the factor of $1.7$ between the L1 peak and the L2 buffet frequency; whether they reflect resolution or lower numerical dissipation is not separated here.
A least-squares fit of the L1 response's surface pressure onto the three certified eigenvectors under the peak captures $99.8\%$ of it, each eigenvector alone $82$ to $88\%$: the response is the interference of a band of damped buffet-cell modes, closer to one another than to the axis.

\begin{table}[!htbp]
\centering
\caption{Spanwise wavelength $\lambda_y$ and outboard phase speed $v_y$ of the surface-pressure cells, from the phase of $\hat p$ on the upper surface; the span band holds $80\%$ of the surface energy.}
\label{tab:crm3M_cells}
\small
\begin{tabular}{llccc}
\toprule
level & structure & span band & $\lambda_y/c_{\mathrm{loc}}$ & $v_y/U_\infty$ \\
\midrule
L2 & unstable eigenvector, $0.378+3.741\,i$ & $0.64$--$0.74$ & $0.61$ & $0.24$ \\
L2 & response, $\omega=3.741$ & $0.64$--$0.74$ & $0.60$ & $0.24$ \\
L1 & response, $\omega=6.4$ & $0.66$--$0.75$ & $0.42$ & $0.28$ \\
L1 & eigenvector, $-0.489+6.129\,i$ & $0.69$--$0.78$ & $0.53$ & $0.33$ \\
L1 & forcing, $\omega=6.4$ & $0.45$--$0.51$ & $0.24$ & $0.21$ \\
L0 & response, $\omega=6.4$ & $0.64$--$0.75$ & $0.43$ & $0.30$ \\
L0 & response, $\omega=7.7$ & $0.64$--$0.74$ & $0.35$ & $0.29$ \\
\bottomrule
\end{tabular}
\end{table}

\subsection{Oscillator or Amplifier}
\label{sec:wavemaker}

\begin{figure}[!htbp]
    \centering
    \includegraphics[width=0.92\textwidth]{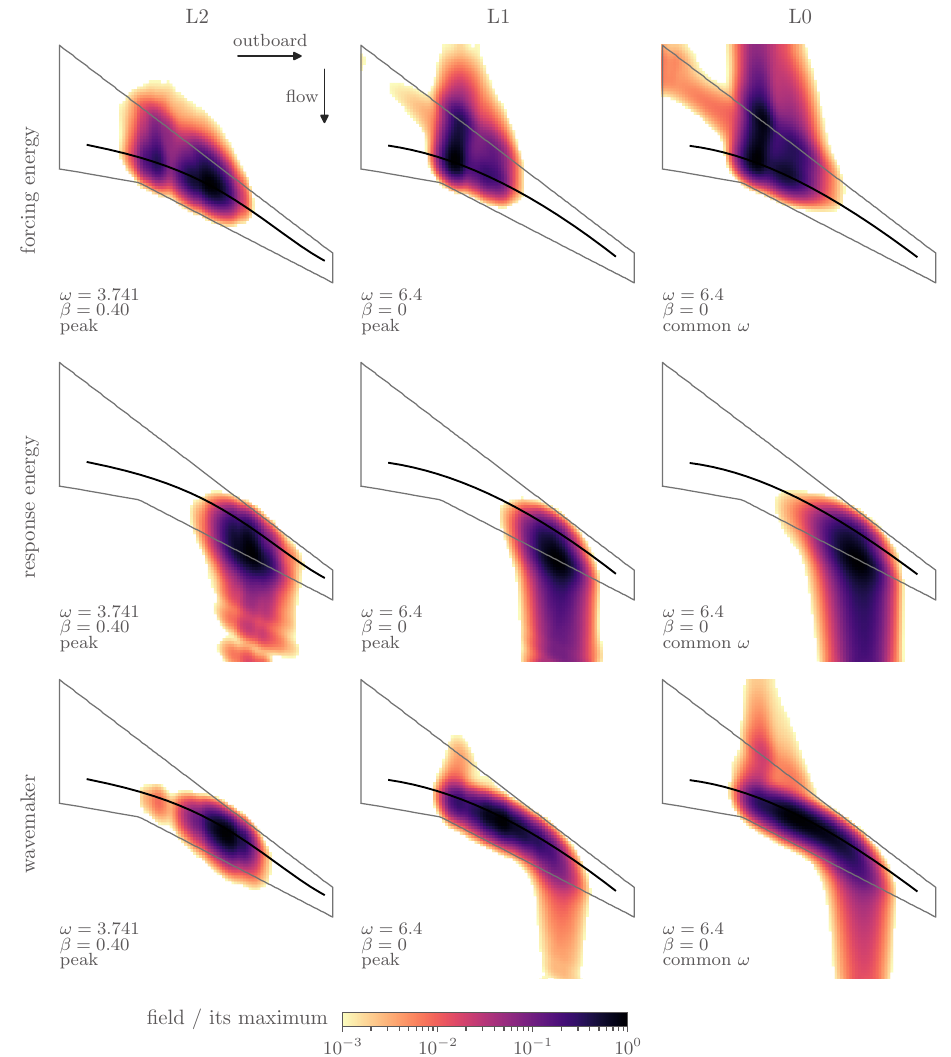}
    \caption{Column-integrated forcing energy (top), response energy (middle) and wavemaker, their pointwise product (bottom), in the volume-weighted Chu norm, on L2, L1 and L0 (left to right), seen from above with the flow from top to bottom and the shock line (black).
    Each map is normalized by its own maximum and drawn on a logarithmic scale, so colours do not compare amplitudes across panels or grids.
    Each panel gives its frequency and contour; L0 is at the common frequency $\omega=6.4$.}
    \label{fig:crm3M_wavemaker}
\end{figure}

The wavemaker (Fig.~\ref{fig:crm3M_wavemaker}) is compact at the shock foot on L2, between $52\%$ and $64\%$ semi-span, where the global-stability wavemaker of the unstable CRM in DLR-TAU also lies~\cite{HoutmanTimme2023}.
At the L2 resonance the optimal forcing is the adjoint eigenvector, so the wavemaker is the structural sensitivity of the unstable mode~\cite{Giannetti2007}.
On L1 it spreads along the swept shock from $38\%$ to $66\%$ semi-span, between the inboard forcing and the outboard response, and L0 keeps it there.
The overlap $\operatorname{Re}\langle\mb{f}_1,\mb{u}_1\rangle_W$ falls from $1.3\times10^{-2}$ on L2 to $1.0\times10^{-3}$ on L1.

Stability decides whether the flow is an oscillator, which buffets without forcing, or an amplifier, which responds only while it is forced~\cite{Huerre1990}.
It does not decide whether the peak belongs to one mode or to several, because non-normality lets one pole reach any height $\kappa_\lambda/\delta$; on L1 it would need $\kappa_\lambda\approx7000$.
The index $\xi\delta$ decides it.
On L1 the central difference of $\ln\sigma_1$ over $\beta=\pm0.05$ gives $\xi=11.9$, six times the $1/\delta=1.9$ of the nearest eigenvalue, so $\xi\delta=6.3$; this is the primary value.
The overlap gives $13.5$ and $\xi\delta=7.2$, higher because the semi-definite Chu weight leaves the $\tilde\nu$ response out of Eq.~\eqref{eq:xi}.
The same three gains give $\delta_{\mathrm{res}}=0.89$ and $m_\beta=10.6$: the gain alone also identifies the peak as collective.
Table~\ref{tab:xidelta} collects the readings.
On L2 the overlap gives $\xi\delta=1.4$, the isolated-pole value $\cos\psi\approx1$ within the non-normal correction of the second unstable pole, $0.20$ away: the peak is the resonance of the unstable mode, and L2 is an oscillator.
On L0 the overlap gives $\xi\delta=4.9$, a lower index than on L1 but still collective; within the shifts searched, L1 and L0 are amplifiers.

\begin{table}[!htbp]
\centering
\caption{Unstable eigenvalues within the shifts searched and the index $\xi\delta$ at the gain peak. On L1 $\xi$ is the central difference of the gain, with the overlap estimate in parentheses; on L2 and L0 it is the overlap estimate.}
\label{tab:xidelta}
\small
\begin{tabular}{llccl}
\toprule
case & unstable eigenvalue & $\delta$ & $\xi\delta$ & reading \\
\midrule
L2, $\omega=3.74$, $\beta=0.40$ & found & $0.022$ & $1.4$ & resonance of the unstable mode: oscillator \\
L1, $\omega=6.4$, $\beta=0$ & none found & $0.53$ & $6.3$ ($7.2$) & collective pseudo-resonance of damped modes \\
L0, $\omega=7.7$, $\beta=0$ & none found & $0.60$ & $4.9$ & collective, lower index \\
\bottomrule
\end{tabular}
\end{table}

A pseudo-resonant peak is fragile: a change $\Delta$ in the damping of every mode multiplies the gain by about $e^{\xi\Delta}$, so the factor of $5.5$ lost from L1 to L0 corresponds to a uniform shift of the band by only $0.13$ to $0.20$.
The fine levels show the pre-onset form of the mechanism that is unstable on the coarse mesh, the regime \citet{Houtman2023} reach at $3.5^\circ$ through incidence and the present grids reach at $4.2^\circ$ through refinement.
The optimal forcing lies inboard ahead of the shock and the response outboard, as cells near $\mathrm{St}=1$.
This separation suggests an open pathway but does not by itself exclude feedback; the amplifier reading rests on the searched spectrum and on the decay of the seeded and pitch-pulse simulations (Section~\ref{sec:stability}).

\section{Computational Cost}
\label{sec:cost}

\begin{figure}[!htbp]
    \centering
    \includegraphics[width=\textwidth]{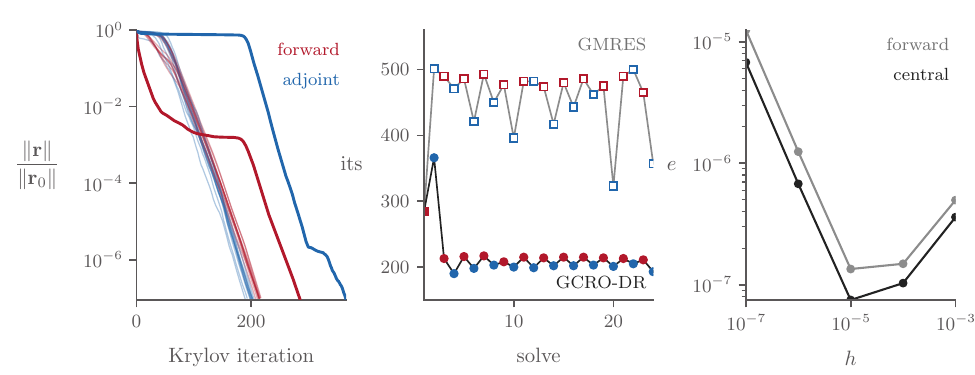}
    \caption{Solver evidence on L1 at $\omega=6.4$, $\beta=0.05$, volume-weighted Chu norm.
    Left: relative residual of every forward (red) and adjoint (blue) solve of one SVD with GCRO-DR, the first pair bold.
    Middle: Krylov iterations of each solve in the order the SVD issues them, GCRO-DR (filled) and GMRES with the same preconditioner (open).
    Right: $e=\|\mathrm{AD}-\mathrm{FD}\|/\|\mathrm{AD}\|$ for a forward and a central finite difference of the residual against the step $h$.}
    \label{fig:crm_solver}
\end{figure}

On one 128-core node a certified L2 frequency costs $5.0$~min matrix-free and $3.6$~min with the Jacobian assembled, nearly all of it 22 inner solves of about $80$ GCRO-DR iterations; the cost is the same on and off resonance.
Matrix-free, a solve costs $1.4$ times as much and uses $31\%$ less memory per rank, and it is the only path on L1 and L0.
Figure~\ref{fig:crm_solver} shows the solves at the L1 peak, $\omega=6.4$ on $\beta=0.05$, in the volume-weighted Chu norm, on 512 ranks (four 128-core nodes) with inner tolerance $10^{-7}$ and SVD tolerance $10^{-5}$.
Every forward and adjoint solve reaches its tolerance without stagnation, the adjoint in $190$ to $205$ iterations against $208$ to $217$ for the forward.
With the same preconditioner, recycling cuts the SVD from $10\,806$ to $5\,198$ iterations and from $3504$ to $1826$~s, a measured speedup of $1.92$ ($48\%$ less time), at the same $\sigma_1$ to $1.6\times10^{-10}$.
On L1 a certified gain costs $34$~min on 512 ranks.
On L0 the $\omega=6.4$ point costs $20.3$~h on 512 ranks, about $10\,400$ rank-hours, with nearly three times the iterations of L1; five L0 sweep points, $\omega=5.2$ to $9.4$, used an SVD tolerance of $10^{-4}$ and a four-vector basis.
Costs here count active ranks, one core each; L0 ran on $2048$ allocated cores, so its allocated cost is four times larger, $41\,600$ core-hours and $C_\omega=98$ to $240$.

\begin{figure}[H]
    \centering
    \includegraphics[width=\textwidth]{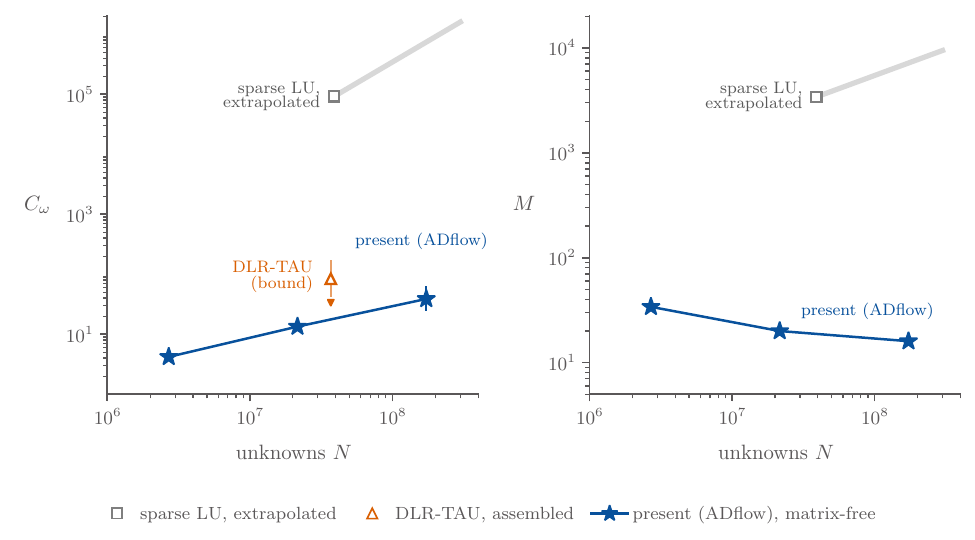}
    \caption{Normalized cost and memory of large three-dimensional resolvent computations against the number of complex unknowns $N$: the sparse factorization extrapolated by \citet{Farghadan2025}, the DLR-TAU CRM resolvent~\cite{Houtman2023,Houtman2022} and the present framework on L2, L1 and L0. Left: core-hours per frequency per $10^{6}$ unknowns, $C_\omega$, counting active ranks; the arrow marks the DLR-TAU upper bound and the L0 bar the range over its points. Right: peak memory per unknown, $M$, in kB. Gray lines: the factorization scalings $N^{2.4}$ and $N^{1.5}$ of \citet{Farghadan2025}.}
    \label{fig:cost_comparison}
\end{figure}

Figure~\ref{fig:cost_comparison} compares two size-free indices across published large-scale resolvent computations: the core-hours per frequency per million complex unknowns, $C_\omega$, and the peak memory per unknown, $M$.
Against the sparse-factorization cost that \citet{Farghadan2025} extrapolate for a jet, a different problem, the framework is two orders of magnitude lower in memory and nearly four in cost.
On L1 its $C_\omega$, $13$, is below the $22$ of RSVD-$\Delta t$~\cite{Farghadan2025} and within the upper bound of $43$ to $164$ that the published DLR-TAU timings allow~\cite{Houtman2023,Houtman2022}.
These are cross-study comparisons on different problems and hardware; the recycling speedup above is the only measured one.
$C_\omega$ rises with size, from $4$ on L2 to $25$--$60$ on L0, because the one-level Schwarz preconditioner needs more iterations as the subdomains multiply.
$M$ falls instead, from $34$~kB on L2 to $16$~kB on L0, as a fixed overhead of about $0.2$--$0.4$~GB per rank is amortized over the method's own $15$~kB per unknown.
A time-domain method, which pays once for a band of frequencies, is cheaper per mode for broadband sweeps of stable flows; the frequency-domain formulation evaluates the resolvent at any shift, including those of unstable operators, and here its cost per frequency was the same on and off resonance.

\section{Conclusions}
\label{sec:conclusions}

A frequency-domain resolvent framework for the RANS equations was presented in which the shifted Jacobian is never assembled.
Automatic differentiation supplies its action and Hermitian transpose, GCRO-DR recycles Krylov vectors across a Lanczos bidiagonalization, the adjoint converges as the forward solve does, and every triplet is certified by a backward error.
Against an extrapolated sparse-factorization estimate for a different problem it is two orders of magnitude lower in memory and nearly four in cost; recycling alone gives a measured $1.92$-fold speedup.

On the CRM wing at $M=0.85$ and $4.2^\circ$ the coarsest grid carries an unstable buffet mode whose gain peak is its resonance.
The Euclidean and volume norms put the optimal forcing into the turbulence variable; the Chu norm removes this artefact.
On the two finer grids no unstable eigenvalue is found within the shifts searched, and the gain peak is a pseudo-resonance of damped modes with the same outboard-travelling buffet cells.
Within that search, refinement thus turns an oscillator into an amplifier of the same buffet-cell structure, and $\xi\delta$ tells such a collective peak from a single-mode resonance.

The grid convergence of the pseudo-resonant gain and of the buffet-band damping remains open.
This sensitivity to resolution motivates future mesh adaptation targeted at the accuracy of resolvent gains and buffet-mode eigenvalues.
A refinement indicator could combine sensitivities of these quantities with estimates of local discretization error, accounting for changes in both the base flow and its linearized operator.
Developing and validating such an indicator would connect the present modal diagnostics to targeted mesh refinement.

Beyond wing buffet, the framework could help study low-frequency unsteadiness in shock-wave/boundary-layer interactions, inlet buzz, and separated shear-layer and wake dynamics.
With appropriate base flows and linearized boundary conditions, these applications would use the same analysis to identify amplified frequencies, receptive forcing regions and coherent response structures.
Adding rigid-body motion inputs and integrated force/moment outputs could support estimates of linear aerodynamic stability derivatives for blunt-body entry vehicles, complementing Bayesian estimation from trajectories~\cite{Kassem2025DynamicStability}.

\begin{samepage}
Further developments include a coarse space for the preconditioner and the time-spectral extension to periodic base flows~\cite{Howell2026}.
Differentiating the certified gain through the SVD~\cite{Kanchi2025} would also let amplification enter design optimization as the leading eigenvalue has~\cite{He2022a,Kanchi2026}.

\end{samepage}

The physical lesson of these certified computations is that similar travelling buffet-cell structures can coexist with different predicted stability and with strongly mesh-sensitive amplification.

\section*{Acknowledgments}
This work used Anvil at Purdue University through ACCESS allocations PHY250265 (Differentiable modal analysis and optimization of fluid dynamical systems) and MCH260003, supported by NSF grants \#2138259, \#2138286, \#2138307, \#2137603 and \#2138296.
The authors used a generative AI assistant to edit the manuscript and to write the scripts that plot the figures from the authors' simulation data; it was also used in developing analysis and post-processing code. The authors checked all of this work and take full responsibility for the content of the paper.

\bibliography{bib/references.bib,bib/claudebib.bib,bib/agi.bib}

\end{document}